\documentclass[11pt]{article}
\usepackage{authblk}
\usepackage[final]{acl}

\usepackage{times}
\usepackage{latexsym}

\usepackage[T1]{fontenc}

\usepackage[utf8]{inputenc}
\usepackage{newunicodechar}
\newunicodechar{，}{,}
\usepackage{microtype}

\usepackage{inconsolata}

\usepackage{graphicx}
\usepackage{booktabs}
\usepackage{multirow}
\usepackage{amsmath}
\usepackage{makecell} 
\usepackage{tabularx}
\usepackage{graphicx}
\usepackage{epstopdf}
\usepackage{svg}
\usepackage{enumitem}  
\usepackage{newunicodechar}
\usepackage{adjustbox}
\usepackage{float}
\usepackage{glossaries}
\newunicodechar{～}{\textasciitilde}
\newcommand{\modelname}{ALPBench}

\title{ALPBench: A Benchmark for Attribution-level Long-term Personal Behavior Understanding}

\author{
 Lu Ren\textsuperscript{1,*},
 Junda She\textsuperscript{1,*},
 Xinchen Luo\textsuperscript{1,*},
 Tao Wang\textsuperscript{1,*},
 Xin Ye\textsuperscript{1},
 Xu Zhang\textsuperscript{1},
 \\
 Muxuan Wang\textsuperscript{1},
 Xiao Yang\textsuperscript{1},
 Chenguang Wang\textsuperscript{1},
 Fei Xie\textsuperscript{1},
 Yiwei Zhou\textsuperscript{1},
 Danjun Wu\textsuperscript{1},
 \\
 Guodong Zhang\textsuperscript{1},
 Yifei Hu\textsuperscript{1},
 Guoying Zheng\textsuperscript{1},
 Shujie Yang\textsuperscript{1},
 Xingmei Wang\textsuperscript{1},
 Shiyao Wang \textsuperscript{1},
 \\
 Yukun Zhou\textsuperscript{1},
 Fan Yang\textsuperscript{1},
 Size Li\textsuperscript{1},
 Kuo Cai\textsuperscript{1},
 Qiang Luo\textsuperscript{1},
 Ruiming Tang\textsuperscript{1,$\dagger$},
 Han Li\textsuperscript{1},
 Kun Gai\textsuperscript{1}
\\
 \textsuperscript{1}KuaiShou Inc., Beijing, China
\\
 \small{
   * Equal contribution
 }
\\
 \small{
   \{renlu05, shejunda, luoxinchen, wangtao37, tangruiming\}@kuaishou.com
 }
}

\begin{document}
\maketitle
\begin{abstract}
Recent advances in large language models (LLMs) have highlighted their potential for personalized recommendation, where accurately capturing user preferences remains a key challenge. Leveraging their strong reasoning and generalization capabilities, LLMs offer new opportunities for modeling long-term user behavior. To systematically evaluate this, we introduce \modelname, a \textbf{Bench}mark for \textbf{A}ttribution-level \textbf{L}ong-term \textbf{P}ersonal Behavior Understanding. Unlike item-focused benchmarks, \modelname\ predicts user-interested attribute combinations, enabling ground-truth evaluation even for newly introduced items. 
It models preferences from long-term historical behaviors rather than users’ explicitly expressed requests, better reflecting enduring interests. 
User histories are represented as natural language sequences, allowing interpretable, reasoning-based personalization. 
\modelname\ enables fine-grained evaluation of personalization by focusing on the prediction of attribute combinations task that remains highly challenging for current LLMs due to the need to capture complex interactions among multiple attributes and reason over long-term user behavior sequences.
The dataset is available at \href{https://huggingface.co/datasets/OpenOneRec/ALPBench}{ALP-Bench on Hugging Face}.
\end{abstract}
\section{Introduction}


Large language models (LLMs)~\cite{comanici2025gemini2.5, kimik2, deepseekv32, gpt4} have achieved remarkable success across diverse general-purpose tasks, exhibiting strong capabilities in language understanding~\cite{bbh, mmlu}, programming~\cite{code, codexglue}, and other complex reasoning scenarios. As these models continue to mature, an increasingly important direction is the shift from general intelligence toward personalized intelligence~\cite{liu2025survey}, where models are expected to adapt their behavior to individual users rather than serving a generic population.

Personalized intelligence is inherently multi-faceted, and can manifest across multiple dimensions, including writing style~\cite{salemi2024lamp}, conversational tone~\cite{salemi2024lamp}, and recommendation~\cite{huang2025RecBench+}. Among these dimensions, recommendation ability plays a particularly critical role in evaluating personalized intelligence. From a practical perspective, recommendation systems lie at the core of many real-world applications and directly influence the user experience. 
When applied to recommendation, LLMs can leverage rich and diverse user interaction histories, providing abundant signals for modeling individual preferences.
Moreover, recommendation outcomes are grounded in observable user behaviors, making them relatively straightforward to verify and evaluate at scale. These properties make recommendation a compelling and well-defined testbed for studying how well LLMs can understand and adapt to individual users. 

Most existing benchmarks~\cite{huang2025RecBench+, tan2025can} for evaluating recommendation capabilities frame the task as predicting a specific item that a user will interact with. 
While this formulation has been effective for traditional recommendation models, it introduces fundamental challenges when used to assess the reasoning and personalization abilities of large language models. 
In practical recommendation systems, item-level predictions are strongly influenced by real-time factors, such as cold-start effects for newly introduced items, item popularity, and platform dynamics.
As a result, the ground-truth item is often influenced by factors that are orthogonal to the user’s stable interests. 
For example, a user who is interested in basketball may be recommended very different videos at different times depending on the NBA schedule, trending events, or platform-wide exposure strategies. 
Accurate prediction requires not only understanding the user’s underlying preferences, but also modeling system-level dynamics and real-time signals. 
However, these real-time factors are neither the focus nor the strength of LLMs, and conflating them with preference modeling can obscure the true personalized reasoning capability of the model.

Therefore, we argue that a more appropriate evaluation should decouple stable user interests from transient system dynamics and focus on the aspects of recommendation that are amenable to reasoning and generalization.
To achieve this goal, we propose the Long-Term Personal Behavior Understanding Bench (\modelname), a benchmark that reformulates recommendation evaluation as an attribute prediction task based on user historical behavior. 
This formulation directly targets the preference modeling component of recommendation, which is stable over time and amenable to reasoning.
Furthermore, the task remains easy to verify, as ground-truth labels are naturally derived from the attributes of items actually consumed by the user. 
Finally, the attribute prediction task can support a wide range of downstream applications.
For example, predicted attribute can be further integrated with external retrieval components, such as lexical indices\cite{formal2021splade}, to match an item for recommendation.

In addition to the novel task design, our benchmark also enables a systematic investigation of long-term user understanding. We construct multiple versions of the task with increasing historical sequence lengths, allowing us to study how models scale with more extensive user histories. 
This setting reveals whether a model can consistently aggregate long-term behavioral signals into coherent preference representations, a capability that is essential for personalized intelligence but remains challenging for current LLMs.


Our contributions are summarized as follows:
\begin{itemize}[leftmargin=*,nosep]
    \item We propose attribute-level user interest prediction as a principled objective for evaluating personalized intelligence in LLMs, explicitly decoupling stable user preferences from transient real-time factors (e.g., popularity and platform dynamics) and focusing on long-term, inferable personalization signals.
    
    \item We introduce a new benchmark for long-sequence personalized intelligence, constructed from real-world industrial user behavior data with long historical contexts, covering eight categories and rich attribute annotations.
    
    \item Through extensive evaluations of state-of-the-art LLMs (including Claude, Qwen, GLM, Kimi, Gemini, GPT, DouBao-seed, Minimax and Deepseek families), we show that existing LLMs, despite strong general capabilities, still struggle to effectively exploit \emph{long behavioral sequences} for personalized user understanding.
\end{itemize}

\section{Related Works}

\subsection{Benchmarks for Long Context}
Early benchmarks for long-context evaluation primarily focused on Needle-in-a-Haystack tests~\cite{yu2025hack, lineedlebench, song2025counting}, assessing a model’s ability to retrieve specific facts from very long sequences. Subsequent benchmarks~\cite{zhang2024bench, li2024loogle} extended this to summarization and reasoning, still assuming all relevant information is fully contained in the context. L-Eval~\cite{leval} and LongBench~\cite{bai2024longbench} aggregate existing datasets (e.g., QASPER~\cite{dasigi2021dataset}, GovReport~\cite{huang2021efficient}) into large-scale long-context evaluation suites covering QA, summarization, retrieval, and coding. DeepSearch~\cite{du2025deepresearch} specifically targets evaluation of Deep Research Agents.

More recent work studies long-context evaluation in interactive dialogue. Some generate sessions exceeding 100k tokens~\cite{maharana2024evaluating, castillo2024beyond}, while benchmarks such as LongMemEval~\cite{wu2024longmemeval} and PersonaMem~\cite{jiang2025personalmem} scale contexts up to nearly 1M tokens using persistent persona-driven conversations.

\subsection{Benchmarking Personal Performance}
\begin{table*}[t]
  \centering
  \small
  \renewcommand{\arraystretch}{1.3}
  \setlength{\tabcolsep}{7pt}
  
  \caption{Detailed statistics of the \textbf{\modelname}. 
  \textbf{Context Token Range (k)} indicates the min/max sequence lengths (in thousands of tokens) for the 3-month (3-M), 6-month (6-M), and 1-year (1-Y) interaction history.  
  \textbf{Avg. \#Item.} denote the average interactions per user.
  \textbf{\#Users} represents the total number of unique users.}
  
  \label{tab:data-statistics-refined}
  \begin{tabular}{l ccc ccc c}
    \toprule
    \multirow{2}{*}{\textbf{Categories}} 
    & \multicolumn{3}{c}{\textbf{Context Token Range (k)}} 
    & \multicolumn{3}{c}{\textbf{Avg. \#Inter.}} 
    & \multirow{2}{*}{\textbf{\#Users}} 
    \\
    
    \cmidrule(lr){2-4} \cmidrule(lr){5-7}
    & \textbf{3-M} & \textbf{6-M} & \textbf{1-Y}
    & \textbf{3-M} & \textbf{6-M} & \textbf{1-Y}
    & \\
    
    \midrule
    Pants      & $0.9 \sim 220.2$ & $0.9 \sim 244.6$ & $2.3 \sim 251.1$ & 339 & 591 & 1049 & 100 \\
    Shoes      & $0.6 \sim 146.5$ & $5.5 \sim 204.3$ & $5.5 \sim 273.3$ & 362 & 652 & 1098 & 100  \\
    Apparel    & $0.7 \sim 158.6$ & $1.8 \sim 189.9$ & $2.1 \sim 274.5$ & 367 & 665 & 1147 & 100 \\
    Snacks     & $0.6 \sim 211.7$ & $5.1 \sim 217.5$ & $5.9 \sim 284.7$ & 429 & 696 & 1102 & 100 \\
    Baijiu     & $0.8 \sim 123.9$ & $0.8 \sim 159.8$ & $1.2 \sim 185.4$ & 194 & 256 & 455 & 100  \\
    Badminton Racket & $0.8 \sim 216.9$ & $0.8 \sim 276.1$ & $1.4 \sim 338.9$ & 285 & 380 & 527 & 100 \\
    Cell Phones
               & $0.8 \sim 186.4$ & $0.8 \sim 186.9$ & $1.0 \sim 187.0$ & 216 & 294 & 487 & 100 \\ 
    Fishing Rods
               & $0.9 \sim 125.4$ & $0.9 \sim 223.9$ & $1.2 \sim 322.5$ & 201 &  284 & 496 & 100 \\
    \bottomrule
  \end{tabular}
\end{table*}

Existing benchmarks for evaluating personalization in LLMs can be roughly grouped by focus.
Some, such as LaMP~\cite{salemi2024lamp, salemi2025lamp}, PDR-Bench~\cite{pdrbench}, PrefEval~\cite{prefeval} and PersonaFeedback~\cite{tao2025personafeedback}, emphasize fine-grained response-level personalization, providing classification or generation tasks, research-focused scenarios, or human-annotated benchmarks for tailoring responses to explicitly specified personas. Others, including PersonaMem~\cite{jiang2025personalmem}, PersonaLLM~\cite{jiangpersonallm}, and PREFDISCO~\cite{PREFDISCO}, evaluating the ability of LLMs to follow or adapt to evolving user personas over multi-round conversations, including dialogues exceeding 100k tokens and multiple topics.

\section{Long-Term Personal Behavior Understanding Benchmark}

\begin{figure*}[htbp]
  \centering
  \includegraphics[width=0.95\textwidth]{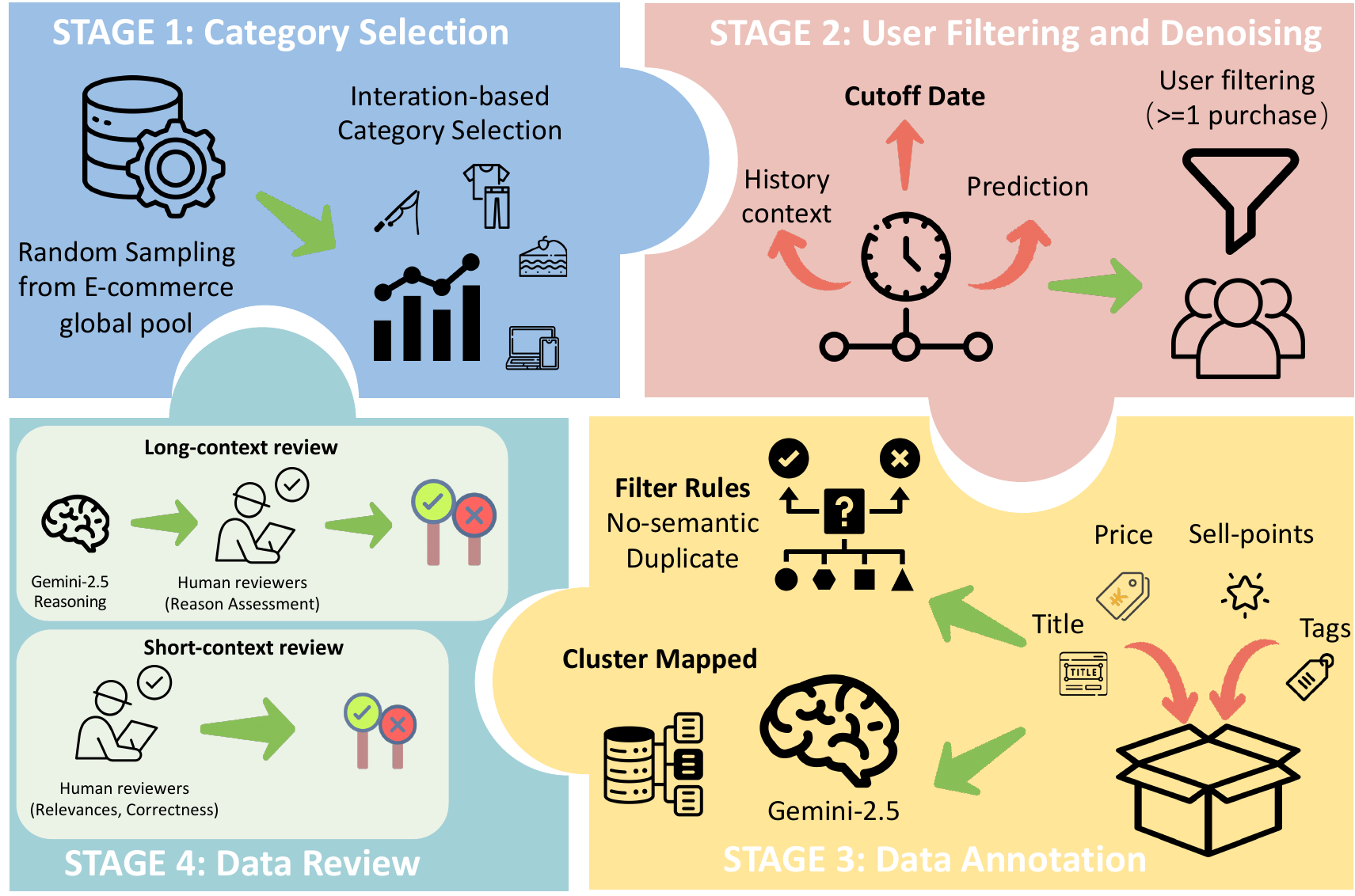} 
  \caption{The complete data collection process of \modelname, including four stages: data collection, data filtering, data cleaning, and data review.}
  \label{fig:process}
\end{figure*}

The Long-Term Personal Behavior Understanding Benchmark (\modelname) is constructed from real-world Chinese e-commerce interaction data collected on the Kuaishou platform and is designed to evaluate large language models in long-context, language-grounded personalization scenarios. Leveraging users' historical item interactions—including clicks, add-to-cart actions, and purchases \modelname\ requires models to predict the attributes of items that users are most likely to purchase next. Unlike setups where models respond to explicit user queries, \modelname\ requires the implicit inference of user preferences from historical behavioral trajectories. All ground-truth labels are derived directly from authentic user interactions, ensuring an objective and verifiable evaluation. To support effective long-context modeling, input contexts are enriched with comprehensive behavioral and item-level features, providing dense signals to enable accurate preference prediction.

\subsection{Overview of \modelname}
\noindent \textbf{Context.}
User historical behavior sequences are constructed from e-commerce interactions, including clicks, add-to-cart actions, and purchases, spanning three temporal horizons: 3, 6, and 12 months. All interactions are ordered chronologically to form a long-range behavioral context.

For each item within the behavioral sequence, we incorporate a multi-faceted set of attributes to enrich the semantic representation:

\begin{itemize}
    \item \textbf{Title.} The original product title, which provides a summary of the item.
    \item \textbf{Selling Points.}  A distilled, platform-curated summary highlighting core functionalities and value propositions. This attribute captures the specific item characteristics most likely to trigger user engagement.
    \item \textbf{Price Tiers.} We categorize price into three levels (low, medium and high), encompassing both the final transaction price for orders and the displayed price during interactions.
\end{itemize}

The statistical metadata of \modelname, including context length, the number of interacted items, and user population, are summarized in Table~\ref{tab:data-statistics-refined}, while the attribute prediction details for the eight sub-tasks are provided in the Appendix~\ref{sec:appendix}.

\noindent \textbf{Formation of task.}
\modelname\ is designed to infer coherent combinations of product attributes that reflect users’ latent preferences. Rather than predicting each attribute independently, it requires models to jointly reason over multiple interdependent attributes to identify a consistent preference profile.

Let $\mathcal{A}_C = \{A_1, A_2, \dots, A_k\}$ denote the set of target attributes for a given category $C$. For each attribute $A_j$, a candidate value set $\mathcal{V}_j$ is provided. The model is required to output a preference profile $\mathbf{y} = \{v_1, v_2, \dots, v_k\}$, where $v_j \in \mathcal{V}_j$ .
This objective can be formalized as selecting the optimal attribute combination $\mathbf{y}^*$ from the joint combinatorial space $\mathcal{V} = \mathcal{V}_1 \times \mathcal{V}_2 \times \dots \times \mathcal{V}_k$:
\begin{equation}
    \hat{\mathbf{y}} = \text{LLM}(S_u, C, \mathcal{A}_C, \{\mathcal{V}_1, \dots, \mathcal{V}_k\}),
\end{equation}
where $S_u$ denotes the user’s long-range behavioral context. This formulation requires large language models to perform reasoning over a attribute space, grounded in personal behavioral histories.

\subsection{Dataset Collection}
Real-world e-commerce user logs are large and heterogeneous, often containing incomplete or inconsistent product metadata. These issues make it difficult to generate accurate evaluation benchmarks. To address this, we design a two-stage data curation pipeline: (1) interaction-based category selection, and (2) user filtering and context denoising. The first stage mitigates metadata inconsistencies by focusing on high-frequency categories, while the second stage removes low-intent actions and anomalous noise to refine the interaction data. The overall process is illustrated in the Figure~\ref{fig:process}.

\noindent \textbf{Interaction-based Category Selection.} 
To alleviate noise introduced by fragmented product metadata, we analyze interaction statistics from the Kuaishou e-commerce pool and select high-frequency categories as prediction targets. These categories yield more reliable behavioral signals, allowing us to suppress label noise from sparse interation actions and achieve a more stable characterization of user preferences.

\noindent \textbf{User Filtering and Denoising.} 
To purify the interaction space from anomalous noise, we leverage the behavioral patterns characteristic of major shopping festivals. We posit that during such periods, purchasing decisions are more deliberate and significantly more reflective of a user's stable, long-term preferences compared to routine browsing. Therefore, we adopt a strict temporal partitioning strategy: We chose the cut-off time which is a shopping festival. The interactions occurring before the cut-off time are treated as the historical context, while subsequent verified purchase events after the cut-off time serve as the ground-truth targets.

We retain only those users who have at least one confirmed purchase record within their historical context. It filters out inactive or low-intent users whose behavioral sequences are too sparse to yield meaningful patterns.

\subsection{Data cleaning}

\noindent \textbf{Context Cleaning.}
To construct a robust representation of the user’s journey, we collect the titles, selling points, and price tiers for each product in the interaction sequence. Integrating these heterogeneous signals yields a holistic semantic representation of individual items. To refine the context, we apply a two-step noise reduction process: (1) filtering out purely functional or non-informative content such as URLs, and (2) removing immediate duplicates to reduce redundancy. In addition, LLM-based semantic checks are applied to identify and remove residual noisy signals.

\noindent \textbf{Label Cleaning.}
Raw attribute values often contain repetitions, inconsistencies, or ambiguous expressions. To address this, we apply an auxiliary semantic normalization process, using a LLM (e.g. Gemini-2.5-Pro) to filter and map synonymous or near-duplicate attribute values to canonical forms. This procedure improves attribute consistency and enhances the verifiability of the resulting data. Examples of attributes filtered by the large model are provided in Appendix~\ref{sec:attr}.

\subsection{Data review}
To ensure the quality and reliability of \modelname, we adopt a carefully structured quality assurance process that integrates human review with model-assisted support mechanisms. The review protocol is outlined below:

Given that most samples in our dataset involve long sequences, the protocol is primarily designed for long-context verification. For cases where fully manual inspection is impractical, we employ a scalable, assisted review approach. An auxiliary model first generates a candidate answer along with an explicit reasoning trace, providing a structured summary of how the task can be solved based on the complete interaction history. Human reviewers then assess whether the reasoning is coherent, logically valid, and sufficiently supported by the underlying context. All final quality judgments are made by human reviewers, with model-generated outputs serving only to support efficient and consistent evaluation. For a few of short-context samples, Manual checks is used to verify the data in terms of both relevance and correctness.
\begin{table*}[t]
\centering
\small
\renewcommand{\arraystretch}{1.4}
\setlength{\tabcolsep}{2pt}
\caption{Performance Comparison of Different LLMs on \modelname\ with 1Y User Interaction Contexts(in \%). Within each LLM family, the best result is highlighted in bold.}
\label{tab:main_results}
\begin{adjustbox}{width=\textwidth} 
\begin{tabular*}{\linewidth}{@{\extracolsep{\fill}}l cccc cccc cccc cccc @{}}
\toprule
\multirow{2}{*}{\textbf{Model}} 
& \multicolumn{2}{c}{\textbf{Pants}} 
& \multicolumn{2}{c}{\textbf{Shoes}} 
& \multicolumn{2}{c}{\textbf{Apparel}} 
& \multicolumn{2}{c}{\textbf{Snacks}}
& \multicolumn{2}{c}{\textbf{Phones}} 
& \multicolumn{2}{c}{\textbf{BaiJiu}} 
& \multicolumn{2}{c}{\textbf{Fish Rods}} 
& \multicolumn{2}{c}{\textbf{Badm. racket}}
\\
\cmidrule(lr){2-3} \cmidrule(lr){4-5} \cmidrule(lr){6-7} \cmidrule(lr){8-9} \cmidrule(lr){10-11} \cmidrule(lr){12-13} \cmidrule(lr){14-15} \cmidrule(lr){16-17}
& F1 & Attr. & F1 & Attr.  & F1 & Attr.  & F1 & Attr.& F1 & Attr. & F1 & Attr.  & F1 & Attr.  & F1 & Attr. \\
\midrule
Qwen3-30B-A3B  & 1.33  & 33.98  & 0.00  & 24.33 & 0.20  & 19.38 & 0.00  & 24.67 & 0.00 & 37.10 & 13.67 & 47.87 & 10.67 & 50.21 & 21.00 & 58.58 \\
Qwen3-235B-A22B  & 2.40  & 39.84 & 1.17  & 25.32 & 1.53  & 23.98 & 0.00  & 21.29  & 0.00 & 41.86  & 16.90 & \textbf{53.54} & 9.33 & 51.90 & 15.80 & 55.90\\
Qwen3-Max-Preview & \textbf{5.89}  & \textbf{55.60}  & \textbf{4.65}  & \textbf{39.36}  & \textbf{1.86}  & \textbf{38.89}  & 0.00  & \textbf{28.13} & \textbf{1.67} & \textbf{42.06} & \textbf{17.80} & 51.46 & \textbf{18.33} & \textbf{55.12} & \textbf{23.67} & \textbf{59.03}\\
\midrule

GLM-4.5  & 3.33  & 42.40  & 2.04  & 30.21  & \textbf{0.17}  & \textbf{29.88}  & \textbf{1.70}  & 18.79 & 0.00 & 37.10  & \textbf{17.63} & 53.35 & \textbf{12.50} & 49.88 & 17.23 & 59.42 \\
GLM-4.6  & \textbf{5.70}  & \textbf{46.90}  & \textbf{6.75}  & \textbf{31.97}  & 0.00  & 29.84  & 0.01  & \textbf{20.12} & 0.00 & \textbf{38.35} & 17.07 & \textbf{58.55} & 9.17 & \textbf{60.89} & \textbf{19.20} & \textbf{62.17} \\

\midrule
Gemini-2.5-Flash  & 4.57 & 49.10 & 3.29 & 33.99 & 0.34 & 34.40 & \textbf{0.29} & \textbf{30.32} & 0.00 & 42.03& 17.53 & 59.92 & 16.81 & 65.21 & \textbf{21.56} & \textbf{63.61}\\

Gemini-2.5-Pro  & \textbf{5.47} & \textbf{56.53} & \textbf{5.65} & \textbf{42.64} & \textbf{1.42} & \textbf{42.66} & 0.22 & 29.93& \textbf{0.33} & \textbf{45.08} & \textbf{19.20} & \textbf{62.16}  & \textbf{18.17} & \textbf{68.48}  & 19.30 & 63.44\\
\midrule

MiniMax-M2  & \textbf{5.26}  & 44.40  & 2.23  & 31.73 & \textbf{1.52}  & 34.50  & 0.00  & 26.90 & 0.00 & 40.71 & \textbf{17.02} & \textbf{56.33} & \textbf{15.17} & 55.32 & \textbf{22.07} & \textbf{58.53} \\
MiniMax-M2.1  & 4.02  & \textbf{54.51} & \textbf{2.92}  & \textbf{38.66} & 0.22  & \textbf{37.55}  & \textbf{0.29}  & \textbf{29.28} & 0.00 & \textbf{42.32} & 16.57 & 54.08 & 13.27 & \textbf{60.35} & 18.77 & 57.74 \\
\midrule

DeepSeek R1 & \textbf{4.00} & 34.88  & 1.07 & 20.89 & 0.00 & 19.61 & 0.00 & \textbf{20.75}& 0.00 & 38.58  & 15.83 & 55.27 & \textbf{16.33} & 53.87 & 16.67 & 54.03 \\
DeepSeek V3.2 & 3.49 & \textbf{37.38}  & \textbf{1.20} & \textbf{24.51} & \textbf{0.33} & \textbf{23.20} & 0.00 & 20.37
& 0.00 & \textbf{40.67}  & \textbf{18.87} & \textbf{63.13} & 14.83 & \textbf{57.90} & \textbf{17.73} & \textbf{59.08} \\
\midrule

GPT-5.2-Chat  & 4.10 & 52.90  & 3.95 & 39.58  & 1.18 & 39.76 & 0.00 & 28.27 & 0.00 & 46.61  & 15.37 & 59.34  & 12.23 & 60.65  & 17.80 & 58.73 \\
\midrule

DouBao  & 6.54 & 56.56 & 4.07 & 39.81  & 0.50 & 40.42 & 0.00 & 31.10 & 0.00 & 45.78 & 20.08 & 65.12 & 18.33 & 65.21 & 25.50 & 65.92\\
\midrule

Claude 4.5 Sonnet  & 5.31 & 46.06  & 4.11 & 32.74  & 1.19 & 29.33  & 0.00 & 17.23 & 0.00 & 43.35 & 18.45 & 63.11  & 13.77 & 59.94 & 19.03 & 59.50\\

\midrule
Kimi & 2.56 & 20.63  & 0.67 & 12.37 & 0.00 & 12.74 & 0.00 & 8.54  & 0.00 & 36.56 & 6.67 & 37.93  & 11.67 & 40.38 & 15.33 & 50.53\\
\bottomrule
\end{tabular*}
\end{adjustbox}
\end{table*}

\section{Experiments}

\subsection{Experiment on \modelname}
\textbf{Experiment Setup.} We evaluate a diverse set of state-of-the-art large language models (LLMs) from multiple research and industrial ecosystems to comprehensively assess their personalized recommendation capabilities under long-context user behavior settings. The evaluated models include Qwen3-30B-A3B-Think-2507, Qwen3-235B-A22B-Instruct-2507~\cite{Qwen/qwen3-30b-a3b-thinking-2507}, and Qwen3-Max-Preview~\cite{qwen3max}; GLM-4.5 and GLM-4.6~\cite{glm}; Gemini-2.5-Pro and Gemini-2.5-Flash; DeepSeek-R1~\cite{deepseekr1} and DeepSeek-V3.2~\cite{deepseekv32}; DouBao-seed-1.8; Claude Sonnet 4.5; GPT-5.2-Chat; as well as Kimi-K2~\cite{kimik2}.

We perform zero-shot evaluations on eight tasks using the selected models, with all inputs presented in a unified format to ensure fair comparison. For completeness, all prompts used in these experiments are provided in Appendix~\ref{sec:prompt_ap}.

\noindent \textbf{Evaluation Metrics.}  
Let $\hat{\mathbf{y}}_i = \{\hat{v}_{i,1}, \dots, \hat{v}_{i,k}\}$ be the predicted preference attribute set for instance $i$, and let $\mathbf{y}_i = \{v_{i,1}, \dots, v_{i,k}\}$ be the corresponding ground-truth attribute set f, where $k$ denotes the number of attributes in each attribute set. Each instance may have multiple valid attribute combinations. We define $\mathcal{D}$ as the set of ground-truth sets and $\mathcal{D}_{\text{pred}}$ as the set of LLM predictions. We evaluate model performance using the following metrics:

\begin{itemize}
\item \textbf{Profile-level Precision.} Measures the proportion of prediction that exactly match the corresponding ground-truth:
\begin{equation}
\text{Precision} = \frac{|\{i \in \mathcal{D}_{\text{pred}} : \hat{\mathbf{y}}_i = \mathbf{y}_i\}|}{|\mathcal{D}_{\text{pred}}|}
\end{equation}

\item \textbf{Profile-level Recall.} Measures the proportion of ground-truth that exactly match the corresponding predictions:
\begin{equation}
\text{Recall} = \frac{|\{i \in \mathcal{D} : \hat{\mathbf{y}}_i = \mathbf{y}_i\}|}{|\mathcal{D}|}
\end{equation}

\item \textbf{Profile-level F1.} The harmonic mean of precision and recall, capturing overall joint correctness:
\begin{equation}
\text{F1}_{\text{profile}} = \frac{2 \cdot \text{Precision} \cdot \text{Recall}}{\text{Precision} + \text{Recall}}
\end{equation}

\item \textbf{Attribute-level F1 (Attr-F1).} Computes the F1 score for each attribute independently and averages across all $k$ attributes to capture fine-grained correctness. $\text{Precision}_j$ and $\text{Recall}_j$ are computed over all instances for attribute $A_j$. This metric reflects the model’s ability to correctly predict each attribute individually.:
\begin{equation}
\text{Attr-F1} = \frac{1}{k} \sum_{j=1}^{k} 
\frac{2 \cdot \text{Precision}_j \cdot \text{Recall}_j}{\text{Precision}_j + \text{Recall}_j},
\end{equation}

\end{itemize}
Profile-level metrics evaluate a model’s ability to generate combinations of attributes, which is relatively more challenging but beneficial for precise item targeting. In contrast, attribute-level metrics assess the model’s ability to generate individual attributes, which is relatively simpler.

\subsection{Main Result of \modelname}




We examine the state-of-the-art LLMs on our benchmark, including Qwen~\cite{Qwen/qwen3-30b-a3b-thinking-2507}, GLM~\cite{glm}, Gemini, MiniMax, and DeepSeek~\cite{deepseekr1, deepseekv32} families.
Table~\ref{tab:main_results} presents the evaluation results of fifteen representative models across eight datasets. 
Overall, the results show clear differences in performance both across model families and evaluation metrics.

Specifically, from the model perspective, we find consistent gaps between models of different sizes within the same family. 
Although a few smaller models occasionally outperform larger ones, the general trend is that larger models significantly outperform smaller ones, demonstrating a strong scaling effect and confirming that our benchmark effectively captures differences in model capability.

From the metric perspective, no single model achieves top performance on all eight datasets. The best-performing model varies by dataset, indicating that different models have distinct inductive biases and are better suited to specific task types or product categories. In addition, performance on profile-level metrics remains low for all models, highlighting the challenge of multi-attribute reasoning, where multiple attributes and their interactions must be considered jointly rather than independently.

\subsection{Performance with Longer Sequence}
\begin{figure*}[htbp]
  \centering
  \includegraphics[width=0.98\textwidth]{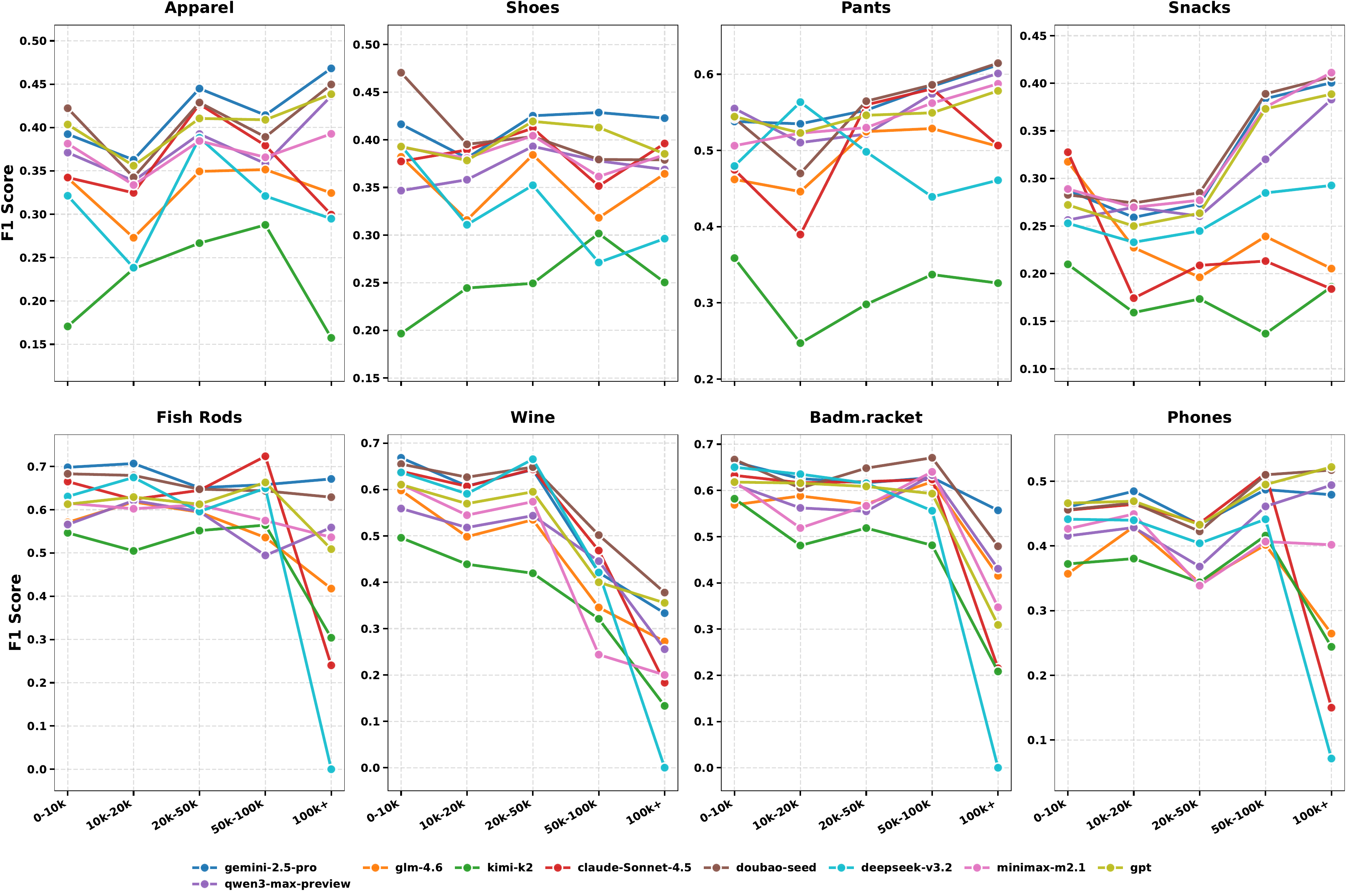} 
  \caption{Performance Comparison of Different LLMs Across Varying Token Lengths.}
  \label{fig:longer}
\end{figure*}

We conduct experiments on the 3M, 6M, and 1Y datasets using DeepSeek-v3.2, Gemini-2.5-Pro, Kimi-K2, Claude-Sonnet-4.5, Doubao-Seed, MiniMax-M2.1, GPT-5.2-Chat, and Qwen-Max. The results are grouped according to input token length and are illustrated in Figure~\ref{fig:longer}. For different categories of tasks, the distinction between models is quite clear and consistent, with models such as DouBao-Seed and Gemini maintaining relatively high performance levels.

Overall, tasks such as Apparel, Shoes, Pants, and Snacks show gradual improvements in performance with increasing context length when using models like Gemini, Qwen, and DouBao. In contrast, models such as Kimi, Claude, and GLM exhibit a rising-then-falling trend, suggesting that their performance initially benefits from longer contexts but deteriorates once the context exceeds their effective capacity. This pattern can be explained by the fact that these categories correspond to high-frequency consumption scenarios: as context length increases, the signal-to-noise ratio (SNR) improves, enabling models to better capture users’ long-term and stable preferences from extended interaction histories.

Non-routine or non-daily-consumption categories, such as Fish Rods, Wine, Badminton Rackets, and Phones, exhibit a distinctly different pattern. Performance in these categories drops sharply as interaction histories grow longer, which we attribute to the Lost in the Middle phenomenon~\cite{liu2024lost}. When the input exceeds a model’s effective context capacity, it struggles to maintain logical consistency over long sequences, leading to a cliff-like decline in performance.

\section{Limitations}

\modelname\ has several limitations that warrant consideration. First, the current benchmark relies exclusively on textual data; richer feature engineering, or the integration of additional modalities—such as images, user interaction signals, or product metadata—could provide higher signal-to-noise ratios and more diverse evaluation scenarios. Incorporating multimodal inputs in future iterations may allow models to better capture complex product attributes and nuanced user behaviors.

Second, although \modelname\ is designed to evaluate model reasoning in long-context personalization settings, its direct integration with practical recommendation systems remains preliminary. Future work could investigate how insights from benchmark evaluations can be translated into improvements in LLM-based recommendation pipelines, including personalized ranking, multi-step decision making and etc. Such efforts would increase the practical utility of \modelname.

Finally, the benchmark currently covers a limited set of e-commerce categories and user behaviors. While these categories provide a meaningful range of personalization scenarios, the findings may not fully generalize to other domains, such as finance, healthcare, or social media, or to tasks requiring different types of reasoning or interaction patterns. Expanding the benchmark to include more diverse product categories, cross-domain datasets, and longer-term user behaviors could strengthen its generalizability and applicability to a wider range of personalization challenges.

\newpage

\bibliography{custom}

@inproceedings{formal2021splade,
  title={SPLADE: Sparse lexical and expansion model for first stage ranking},
  author={Formal, Thibault and Piwowarski, Benjamin and Clinchant, St{\'e}phane},
  booktitle={Proceedings of the 44th International ACM SIGIR Conference on Research and Development in Information Retrieval},
  pages={2288--2292},
  year={2021}
}

@article{mmlu,
  title={Measuring massive multitask language understanding},
  author={Hendrycks, Dan and Burns, Collin and Basart, Steven and Zou, Andy and Mazeika, Mantas and Song, Dawn and Steinhardt, Jacob},
  journal={arXiv preprint arXiv:2009.03300},
  year={2020}
}

@inproceedings{bbh,
  title={Challenging big-bench tasks and whether chain-of-thought can solve them},
  author={Suzgun, Mirac and Scales, Nathan and Sch{\"a}rli, Nathanael and Gehrmann, Sebastian and Tay, Yi and Chung, Hyung Won and Chowdhery, Aakanksha and Le, Quoc and Chi, Ed and Zhou, Denny and others},
  booktitle={Findings of the Association for Computational Linguistics: ACL 2023},
  pages={13003--13051},
  year={2023}
}

@article{codexglue,
  title={Codexglue: A machine learning benchmark dataset for code understanding and generation},
  author={Lu, Shuai and Guo, Daya and Ren, Shuo and Huang, Junjie and Svyatkovskiy, Alexey and Blanco, Ambrosio and Clement, Colin and Drain, Dawn and Jiang, Daxin and Tang, Duyu and others},
  journal={arXiv preprint arXiv:2102.04664},
  year={2021}
}

@article{code,
  title={Evaluating large language models trained on code},
  author={Chen, Mark},
  journal={arXiv preprint arXiv:2107.03374},
  year={2021}
}

@article{tan2025can,
  title={Can Large Language Models Understand Preferences in Personalized Recommendation?},
  author={Tan, Zhaoxuan and Zeng, Zinan and Zeng, Qingkai and Wu, Zhenyu and Liu, Zheyuan and Mo, Fengran and Jiang, Meng},
  journal={arXiv preprint arXiv:2501.13391},
  year={2025}
}

@article{liu2024lost,
  title={Lost in the middle: How language models use long contexts},
  author={Liu, Nelson F and Lin, Kevin and Hewitt, John and Paranjape, Ashwin and Bevilacqua, Michele and Petroni, Fabio and Liang, Percy},
  journal={Transactions of the association for computational linguistics},
  volume={12},
  pages={157--173},
  year={2024}
}

@article{liu2025survey,
  title={A survey of personalized large language models: Progress and future directions},
  author={Liu, Jiahong and Qiu, Zexuan and Li, Zhongyang and Dai, Quanyu and Yu, Wenhao and Zhu, Jieming and Hu, Minda and Yang, Menglin and Chua, Tat-Seng and King, Irwin},
  journal={arXiv preprint arXiv:2502.11528},
  year={2025}
}

@inproceedings{bai2024longbench,
  title={Longbench: A bilingual, multitask benchmark for long context understanding},
  author={Bai, Yushi and Lv, Xin and Zhang, Jiajie and Lyu, Hongchang and Tang, Jiankai and Huang, Zhidian and Du, Zhengxiao and Liu, Xiao and Zeng, Aohan and Hou, Lei and others},
  booktitle={Proceedings of the 62nd annual meeting of the association for computational linguistics (volume 1: Long papers)},
  pages={3119--3137},
  year={2024}
}

@article{huang2021efficient,
  title={Efficient attentions for long document summarization},
  author={Huang, Luyang and Cao, Shuyang and Parulian, Nikolaus and Ji, Heng and Wang, Lu},
  journal={arXiv preprint arXiv:2104.02112},
  year={2021}
}

@article{dasigi2021dataset,
  title={A dataset of information-seeking questions and answers anchored in research papers},
  author={Dasigi, Pradeep and Lo, Kyle and Beltagy, Iz and Cohan, Arman and Smith, Noah A and Gardner, Matt},
  journal={arXiv preprint arXiv:2105.03011},
  year={2021}
}

@article{huang2025RecBench+,
  title={Towards Next-Generation Recommender Systems: A Benchmark for Personalized Recommendation Assistant with LLMs},
  author={Huang, Jiani and Wang, Shijie and Ning, Liang-bo and Fan, Wenqi and Wang, Shuaiqiang and Yin, Dawei and Li, Qing},
  journal={arXiv preprint arXiv:2503.09382},
  year={2025}
}

@inproceedings{salemi2024lamp,
  title={Lamp: When large language models meet personalization},
  author={Salemi, Alireza and Mysore, Sheshera and Bendersky, Michael and Zamani, Hamed},
  booktitle={Proceedings of the 62nd Annual Meeting of the Association for Computational Linguistics},
  pages={7370--7392},
  year={2024}
}

@article{salemi2025lamp,
  title={LaMP-QA: A Benchmark for Personalized Long-form Question Answering},
  author={Salemi, Alireza and Zamani, Hamed},
  journal={arXiv preprint arXiv:2506.00137},
  year={2025}
}

@article{maharana2024evaluating,
  title={Evaluating very long-term conversational memory of llm agents},
  author={Maharana, Adyasha and Lee, Dong-Ho and Tulyakov, Sergey and Bansal, Mohit and Barbieri, Francesco and Fang, Yuwei},
  journal={arXiv preprint arXiv:2402.17753},
  year={2024}
}

@article{castillo2024beyond,
  title={Beyond prompts: Dynamic conversational benchmarking of large language models},
  author={Castillo-Bolado, David and Davidson, Joseph and Gray, Finlay and Rosa, Marek},
  journal={Advances in Neural Information Processing Systems},
  volume={37},
  pages={42528--42565},
  year={2024}
}

@article{jiangpersonallm,
  title={PersonaLLM: investigating the ability of large language models to express personality traits; 2024},
  author={Jiang, Hang and Zhang, Xiajie and Cao, Xubo and Breazeal, Cynthia and Roy, Deb and Kabbara, Jad},
  journal={URL https://api. semanticscholar. org/CorpusID},
  volume={268032940}
}

@inproceedings{leval,
  title={L-eval: Instituting standardized evaluation for long context language models},
  author={An, Chenxin and Gong, Shansan and Zhong, Ming and Zhao, Xingjian and Li, Mukai and Zhang, Jun and Kong, Lingpeng and Qiu, Xipeng},
  booktitle={Proceedings of the 62nd Annual Meeting of the Association for Computational Linguistics},
  pages={14388--14411},
  year={2024}
}

@article{wu2024longmemeval,
  title={Longmemeval: Benchmarking chat assistants on long-term interactive memory},
  author={Wu, Di and Wang, Hongwei and Yu, Wenhao and Zhang, Yuwei and Chang, Kai-Wei and Yu, Dong},
  journal={arXiv preprint arXiv:2410.10813},
  year={2024}
}

@article{jiang2025personalmem,
  title={Know me, respond to me: Benchmarking llms for dynamic user profiling and personalized responses at scale},
  author={Jiang, Bowen and Hao, Zhuoqun and Cho, Young-Min and Li, Bryan and Yuan, Yuan and Chen, Sihao and Ungar, Lyle and Taylor, Camillo J and Roth, Dan},
  journal={arXiv preprint arXiv:2504.14225},
  year={2025}
}

@inproceedings{yu2025hack,
  title={Sequential-niah: A needle-in-a-haystack benchmark for extracting sequential needles from long contexts},
  author={Yu, Yifei and Zhang, Qian-Wen and Qiao, Lingfeng and Yin, Di and Li, Fang and Wang, Jie and Xi, Chen Zeng and Zheng, Suncong and Liang, Xiaolong and Sun, Xing},
  booktitle={Proceedings of the 2025 Conference on Empirical Methods in Natural Language Processing},
  pages={29438--29456},
  year={2025}
}

@article{lineedlebench,
  title={NeedleBench: Evaluating LLM Retrieval and Reasoning Across Varying Information Densities},
  author={Li, Mo and Zhang, Songyang and Zhang, Taolin and Duan, Haodong and Liu, Yunxin and Chen, Kai},
  journal={Transactions on Machine Learning Research}
}

@inproceedings{song2025counting,
  title={Counting-stars: A multi-evidence, position-aware, and scalable benchmark for evaluating long-context large language models},
  author={Song, Mingyang and Zheng, Mao and Luo, Xuan},
  booktitle={Proceedings of the 31st International Conference on Computational Linguistics},
  pages={3753--3763},
  year={2025}
}

@article{pdrbench,
  title={Towards personalized deep research: Benchmarks and evaluations},
  author={Liang, Yuan and Li, Jiaxian and Wang, Yuqing and Wang, Piaohong and Tian, Motong and Liu, Pai and Qiao, Shuofei and Fang, Runnan and Zhu, He and Zhang, Ge and others},
  journal={arXiv preprint arXiv:2509.25106},
  year={2025}
}

@article{PREFDISCO,
  title={Personalized Reasoning: Just-in-Time Personalization and Why LLMs Fail at It},
  author={Li, Shuyue Stella and Bose, Avinandan and Brahman, Faeze and Du, Simon Shaolei and Koh, Pang Wei and Fazel, Maryam and Tsvetkov, Yulia},
  journal={arXiv preprint arXiv:2510.00177},
  year={2025}
}

@article{tao2025personafeedback,
  title={PersonaFeedback: A Large-scale Human-annotated Benchmark For Personalization},
  author={Tao, Meiling and Zhu, Chenghao and Ding, Dongyi and Wang, Tiannan and Jiang, Yuchen Eleanor and Zhou, Wangchunshu},
  journal={arXiv preprint arXiv:2506.12915},
  year={2025}
}

@article{prefeval,
  title={Do llms recognize your preferences? evaluating personalized preference following in llms},
  author={Zhao, Siyan and Hong, Mingyi and Liu, Yang and Hazarika, Devamanyu and Lin, Kaixiang},
  journal={arXiv preprint arXiv:2502.09597},
  year={2025}
}

@article{du2025deepresearch,
  title={DeepResearch Bench: A Comprehensive Benchmark for Deep Research Agents},
  author={Du, Mingxuan and Xu, Benfeng and Zhu, Chiwei and Wang, Xiaorui and Mao, Zhendong},
  journal={arXiv preprint arXiv:2506.11763},
  year={2025}
}

@inproceedings{zhang2024bench,
  title={{$\infty$} Bench: Extending long context evaluation beyond 100k tokens},
  author={Zhang, Xinrong and Chen, Yingfa and Hu, Shengding and Xu, Zihang and Chen, Junhao and Hao, Moo and Han, Xu and Thai, Zhen and Wang, Shuo and Liu, Zhiyuan and others},
  booktitle={Proceedings of the 62nd Annual Meeting of the Association for Computational Linguistics},
  pages={15262--15277},
  year={2024}
}

@inproceedings{li2024loogle,
  title={Loogle: Can long-context language models understand long contexts?},
  author={Li, Jiaqi and Wang, Mengmeng and Zheng, Zilong and Zhang, Muhan},
  booktitle={Proceedings of the 62nd Annual Meeting of the Association for Computational Linguistics},
  pages={16304--16333},
  year={2024}
}

@article{comanici2025gemini2.5,
  title={Gemini 2.5: Pushing the frontier with advanced reasoning, multimodality, long context, and next generation agentic capabilities},
  author={Comanici, Gheorghe and Bieber, Eric and Schaekermann, Mike and Pasupat, Ice and Sachdeva, Noveen and Dhillon, Inderjit and Blistein, Marcel and Ram, Ori and Zhang, Dan and Rosen, Evan and others},
  journal={arXiv preprint arXiv:2507.06261},
  year={2025}
}

@misc{Qwen/qwen3-30b-a3b-thinking-2507,
      title={Qwen3 Technical Report}, 
      author={Qwen Team},
      year={2025},
      eprint={2505.09388},
      archivePrefix={arXiv},
      primaryClass={cs.CL},
      url={https://arxiv.org/abs/2505.09388}, 
}

@misc{qwen3max,
    title = {Qwen3-Max: Just Scale it},
    author = {Qwen Team},
    month = {September},
    year = {2025}
}

@article{glm,
  title={Glm-4.5: Agentic, reasoning, and coding (arc) foundation models},
  author={Zeng, Aohan and Lv, Xin and Zheng, Qinkai and Hou, Zhenyu and Chen, Bin and Xie, Chengxing and Wang, Cunxiang and Yin, Da and Zeng, Hao and Zhang, Jiajie and others},
  journal={arXiv preprint arXiv:2508.06471},
  year={2025}
}

@article{deepseekv32,
  title={Deepseek-v3. 2: Pushing the frontier of open large language models},
  author={Liu, Aixin and Mei, Aoxue and Lin, Bangcai and Xue, Bing and Wang, Bingxuan and Xu, Bingzheng and Wu, Bochao and Zhang, Bowei and Lin, Chaofan and Dong, Chen and others},
  journal={arXiv preprint arXiv:2512.02556},
  year={2025}
}

@article{deepseekr1,
  title={Deepseek-r1: Incentivizing reasoning capability in llms via reinforcement learning},
  author={Guo, Daya and Yang, Dejian and Zhang, Haowei and Song, Junxiao and Zhang, Ruoyu and Xu, Runxin and Zhu, Qihao and Ma, Shirong and Wang, Peiyi and Bi, Xiao and others},
  journal={arXiv preprint arXiv:2501.12948},
  year={2025}
}

@article{kimik2,
  title={Kimi k2: Open agentic intelligence},
  author={Team, Kimi and Bai, Yifan and Bao, Yiping and Chen, Guanduo and Chen, Jiahao and Chen, Ningxin and Chen, Ruijue and Chen, Yanru and Chen, Yuankun and Chen, Yutian and others},
  journal={arXiv preprint arXiv:2507.20534},
  year={2025}
}

@article{gpt4,
  title={Gpt-4 technical report},
  author={Achiam, Josh and Adler, Steven and Agarwal, Sandhini and Ahmad, Lama and Akkaya, Ilge and Aleman, Florencia Leoni and Almeida, Diogo and Altenschmidt, Janko and Altman, Sam and Anadkat, Shyamal and others},
  journal={arXiv preprint arXiv:2303.08774},
  year={2023}
}

\appendix

\newpage
\onecolumn
\section{Details of Data Statics}
\label{sec:appendix}

\subsection{Multi-dimensional Label Space}
Table~\ref{tab:predictable-tags} presents the structured attribute space defined in \modelname. For each product category, we specify a set of domain-specific attributes that form the prediction targets for corresponding sub-tasks. These fine-grained attribute tags, when considered in combination, not only enable a precise evaluation of a model’s ability to infer nuanced user preferences within combinatorial attribute spaces, but also facilitate more accurate and effective product positioning, reflecting the model’s capacity for multi-attribute reasoning and personalization.

\begin{table}[H]
    \centering
    \small
    \renewcommand{\arraystretch}{1.3}
    \setlength{\tabcolsep}{10pt}

    \caption{Predictable Attributes in \modelname.}
    \label{tab:predictable-tags}

    \begin{tabularx}{0.95\textwidth}{l X}
        \toprule
        \textbf{Category} & \textbf{Target Attributes (Predictable Tags)} \\
        \midrule
        Shoes & Price Range, Category, Closure Type, Toe Style\\ 
        Apparel & Price Range, Fit, Thickness, Category, Sleeve Length \\ 
        Pants & Price Range, Fit, Category, Waist Type, Pant Length \\
        Snacks & Price Range, Sugar Content, Flavor, Storage, Category and etc. \\
        Phone & Price Range, Brand, Wireless Charging Support, Battery Capacity, Screen Size, Rear Camera Setup, NFC Support, Charging Power  \\
        Wine & Price Range, Brand, Alcohol Content, Packaging \\
        Fish Rods & Price Range, Brand, Material, Rod Action \\
        Badminton Racket & Price Range, Brand, Material, Level \\
        \bottomrule
    \end{tabularx}
\end{table}
\label{sec:attr}

\subsection{Example Prompt for the Data}
\label{sec:prompt_ap}
\begin{figure}[H]
\centering
\begin{minipage}{0.48\linewidth}
  \centering
  \includegraphics[width=\linewidth]{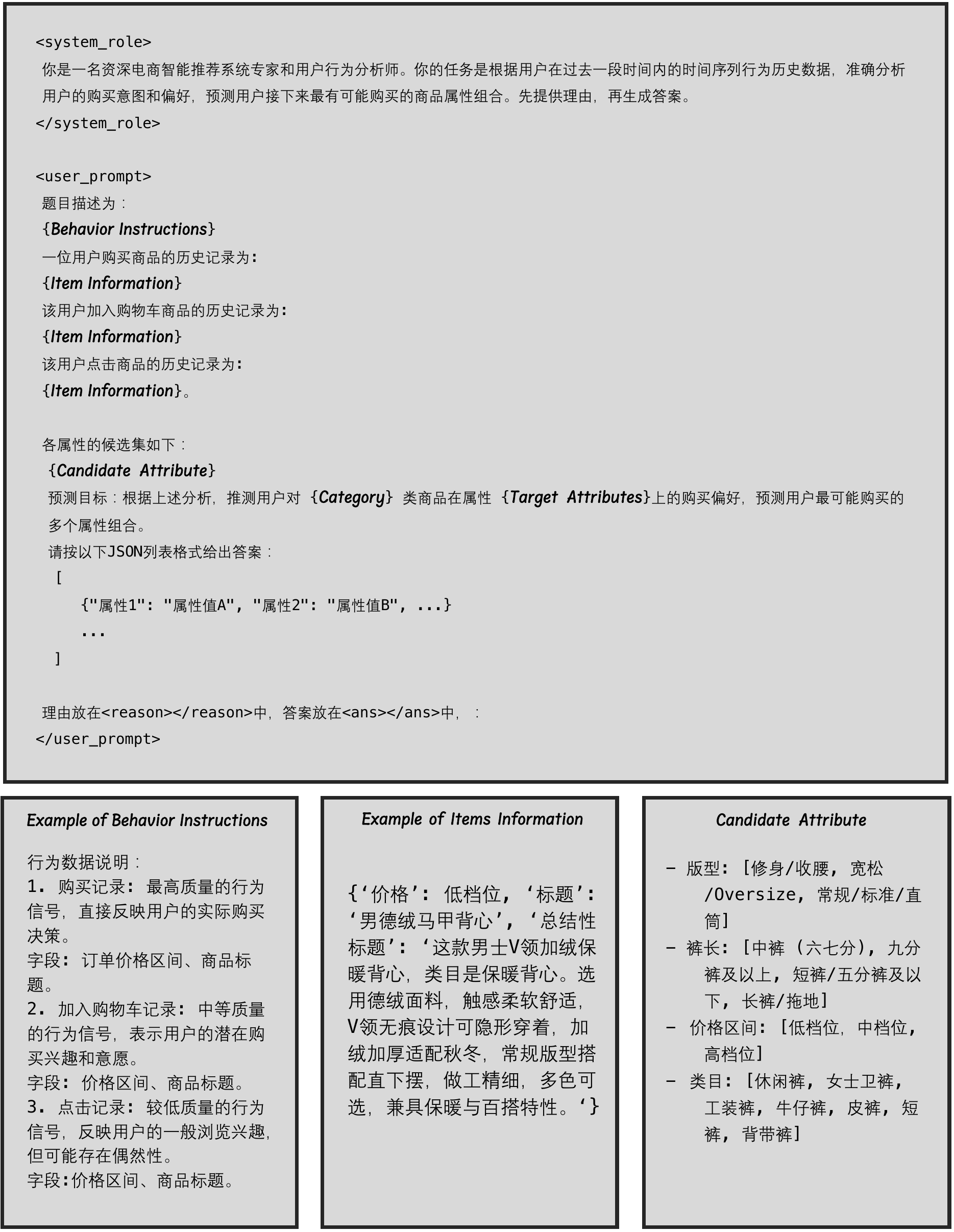}
  \caption{Evaluation prompt template.}
  \label{fig:example-1}
\end{minipage}
\hfill
\begin{minipage}{0.48\linewidth}
  \centering
  \includegraphics[width=\linewidth]{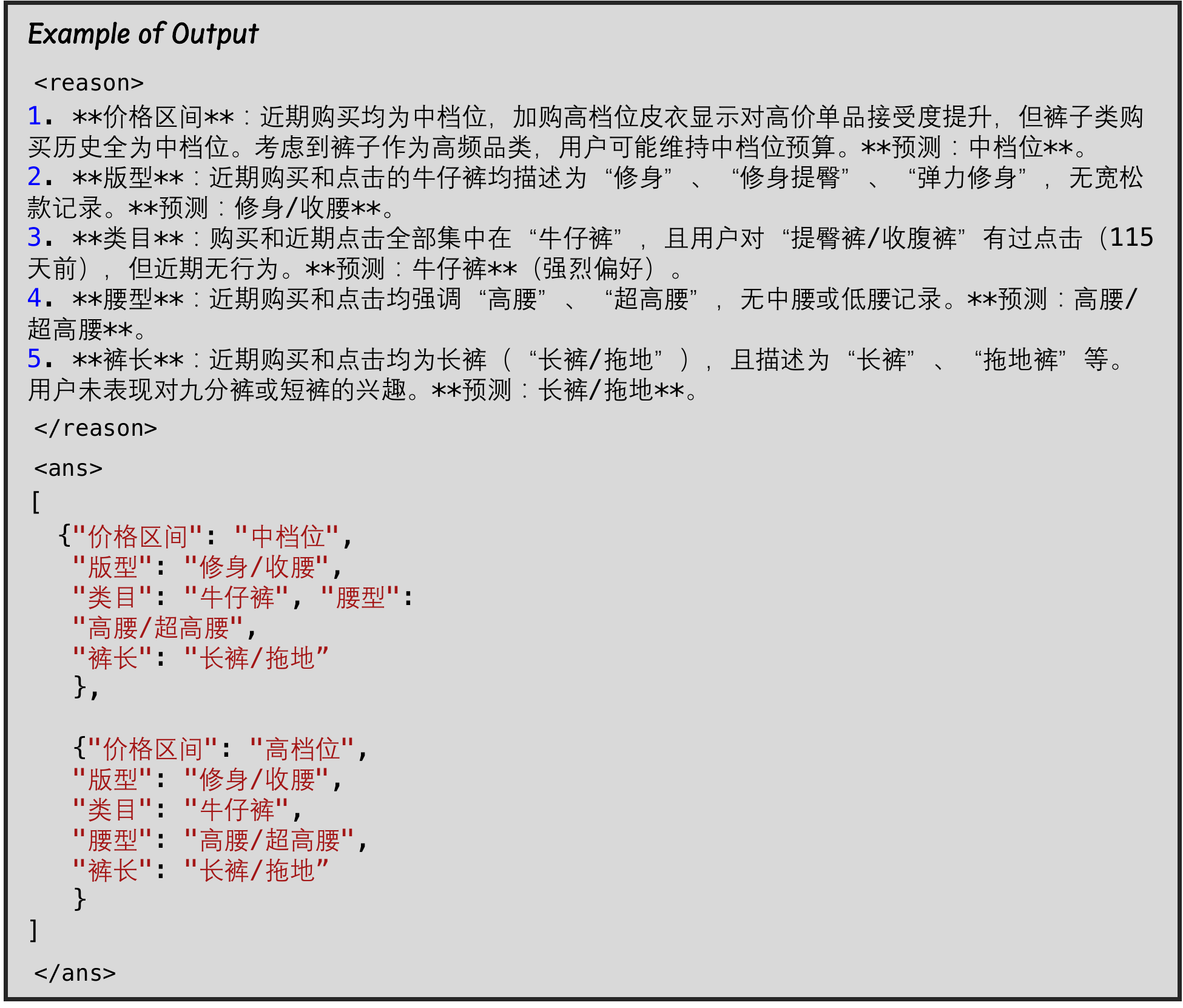}
  \caption{Example of output.}
  \label{fig:example-3}
\end{minipage}
\end{figure}

In this section, we introduce the customized prompt design employed in \modelname. Figures~\ref{fig:example-1} and~\ref{fig:example-2} illustrate the evaluation prompt templates, while Figure~\ref{fig:example-3} presents examples of the model outputs.

\section{Details of Data Collection}

\begin{figure}[H]
  \centering
  \includegraphics[width=0.8\textwidth]{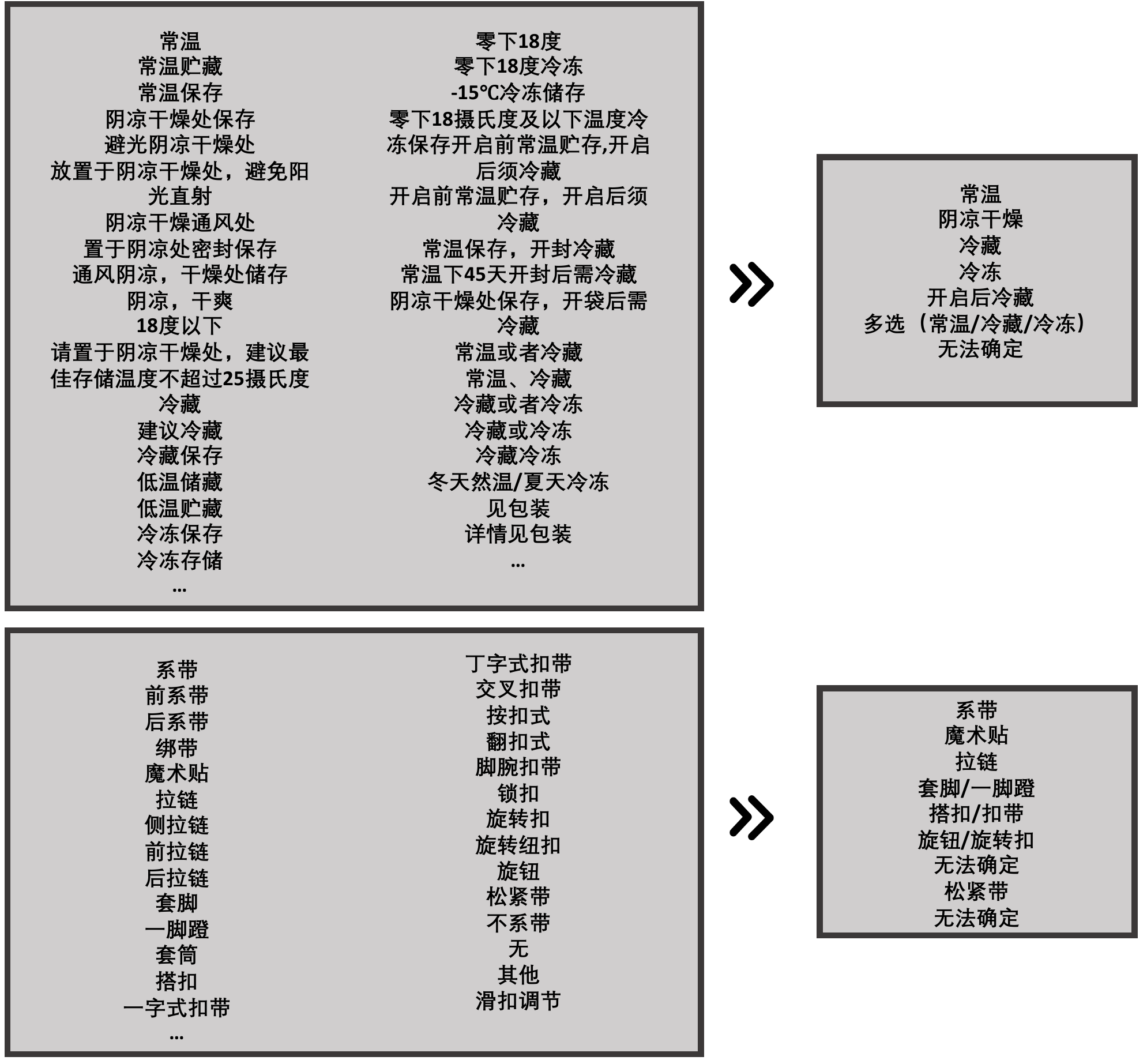} 
  \caption{Attribute Tag Classification Example with Gemini.}
  \label{fig:classification}
\end{figure}

We employ Gemini-2.5-Pro to filter redundant, ambiguous, or incorrectly labeled attributes, a step that is critical for ensuring the benchmark’s reliability and the accuracy of subsequent predictions. Simultaneously, we perform attribute clustering, consolidating numerous semantically similar but non-standardized attribute values into a unified canonical space. Representative examples of this process are illustrated in the Figure~\ref{fig:classification}.

\end{document}